# Lifelog Patterns Analyzation using Graph Embedding based on Deep Neural Network

Wonsup Shin, Tae-Young Kim, and Sung-Bae Cho

Department of Computer Science, Yonsei University, Seoul, Republic of Korea
{wsshin2013, taeyoungkim, sbcho}@yonsei.ac.kr

**Abstract.** Recently, as the spread of smart devices increases, the amount of data collected through sensors is increasing. A lifelog is a kind of big data to analyze behavior patterns in the daily life of individuals collected from various smart devices. However, sensor data is a low-level signal that makes it difficult for humans to recognize the situation directly and cannot express relations clearly. It is also difficult to identify the daily behavior pattern because it records heterogeneous data by various sensors. In this paper, we propose a method to define a graph structure with node and edge and to extract the daily behavior pattern from the generated lifelog graph. We use the graph convolution method to embeds the lifelog graph and maps it to low dimension. The graph convolution layer improves the expressive power of the daily behavior pattern by implanting the lifelog graph in the non-Euclidean space and learns the patterns of graphs. Experimental results show that the proposed method automatically extracts meaningful user patterns from UbiqLog dataset. In addition, we confirm the usefulness by comparing our method with existing methods to evaluate performance.

## 1 Introduction

Recently, as the use of smart devices has expanded, information technology for recording the daily life of individuals in the ubiquitous environment has been rapidly developing [1]. Using various sensors built in smart devices, various kinds of signals can be collected to recognize various situations of users [2]. In addition, data collected from smart devices can be automatically stored to retrieve and utilize past information [3]. However, as the amount of data becomes wider, it is difficult to establish an efficient storage structure considering scalability [4]. Also, it is difficult to extract daily behavior patterns as the data becomes complicated [5].

In this paper, we propose a graph pattern analyzation framework. It constructs graph from lifelogs and pattern extraction method for analyzation. We define the nodes of the graph structure using the context awareness extracted from the input signal of the smart device and connect the edges to express the semantic relation between the nodes clearly. We also propose a graph convolution method that extracts the daily behavior pattern from the generated lifelog graph. The graph convolution method embeds the lifelog graph and maps it to a lower dimension. The embedded lifelog graph is useful for finding frequent user patterns. We also use a novel adversarial auto-encoder model to generate the non-Euclidean feature space. To verify the usefulness of the proposed method, we use UbiqLog (smartphone lifelogging) dataset provided by UCI machine learning repository. The UbiqLog dataset was collected by 35 smartphone users for two months and used the smartphone lifelogging tool. Figure 1 shows the life log data collected over time.

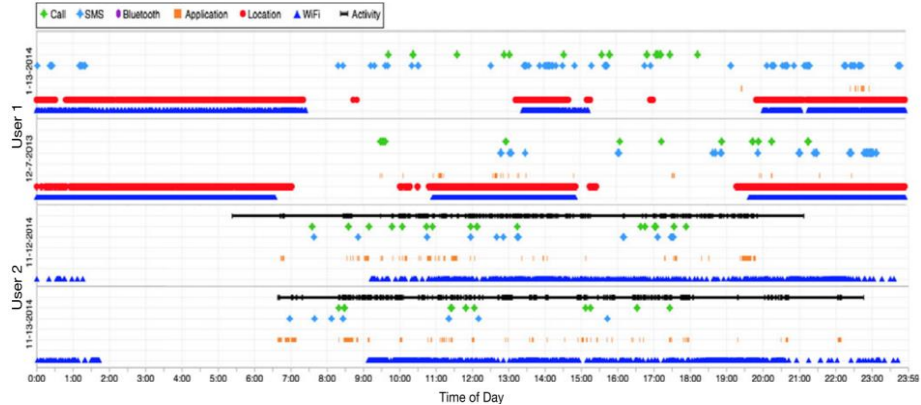

**Figure 1.** UbiqLog data over time

The rest of this paper is organized as follows. We discuss the related work on life log pattern extraction in Section 2. The proposed model is introduced in Section 3. We present the experiments and performance comparison with other methods in Section 4. We conclude this paper with remarks on future work in Section 5.

## 2 Related Work

### 2.1 Previous works to extract patterns from the lifelog

As shown in Table 1, many researchers have studied various storage structures to extract patterns in the lifelog. The lifelog storage structure is typically divided into a linear structure that is stored over time and a graph structure that includes the relationship between data.

This section outlines a variety of previous studies to compare with the proposed method. To illustrate the motivation of the proposed method, we introduce previous research that uses lifelog data.

A. Shimojo et al. used a MySQL database to store food photographs and user weight data collected on Flickr [6]. They proposed a platform that can visualize the eating habit pattern using the SQL query in the constructed database. J. Gemmell et al. used a relational database to record the user's position, calorie, and heart rate [8]. They proposed a platform that can efficiently retrieve information from a built-in lifelog database. These methods can manage the collected life logs by dividing them into tabular form, so that the redundancy can be minimized while maintaining the consistency of the data.

However, when stored in table form, semantic reasoning between data is impossible and only formal queries are possible.

H. Kim et al. Constructed an ontology graph to analyze the data collected through the health tracking system [9]. They classify the data using an ontology-based network and infer the meaning. H. Xu built a semantic network using data collected from smartphones [10]. They use the semantic relationship of the network to shorten the search time and increase the memory usage efficiency. This method can combine and structure lifelogs of various formats and can easily express the relationship and dependency between data. However, the larger the size of the data, the more complex the reasoning becomes. It is also difficult to find patterns in higher dimensional structures.

**Table 1.** Related works to extract patterns from the lifelog

| Category | Author | Year | Description |
|---|---|---|---|
| Lifelog linear structure | A. Shimojo [6] | 2011 | User eating and visualization platform using MySQL |
| | W. V. Sujansky [7] | 2010 | Check patient's health information through SQL query |
| | J. Gemmell [8] | 2006 | SQL-based personal everyday recording platform |
| Lifelog graph structure | H. H. Kim [9] | 2015 | Ontology graph representing biometric data |
| | H. Xu [10] | 2013 | Graph search system utilizing semantic relations |
| | S. Lee [11] | 2010 | Ontology-based life log storage system |

## 3 The Proposed Method

### 3.1 Overview of the Proposed Framework

This paper proposes a framework for analyzing patterns of lifelog data such as cell phone logs. The overall structure of the proposed framework is shown in Fig 2. First, the proposed method expresses log data as a semantic network. This process makes it possible to efficiently express the semantic information of the log data. Second, we embed each network into latent space by using a graph convolution-based deep model. Through this, users' patterns can be expressed on latent space. Finally, we analyze the patterns of each users or groups by using learned encoders.

### 3.2 Graph Generation Process

We set the semantics by applying the theory of General Social Survey, ConceptNet5, and Speech act theory to create a semantic network structure. Table 2 shows the semantics of connections for each node type. Our network consists of three types of nodes through four semantics. The source type nodes reflect actual log data. These store information when the event occurs. The sensor type node reflects the event type. For example, if a user calls two people, there are source nodes for each phone number, which are connected to the 'call' sensor node. The time node reflects the time at which the event occurs. Also, these exist equally for each graph, regardless of the data. Through this, it is possible to represent events occurring in different graphs at the same time flow, and to reduce variation between graphs. Figure 3 shows an example of a network that can be created.

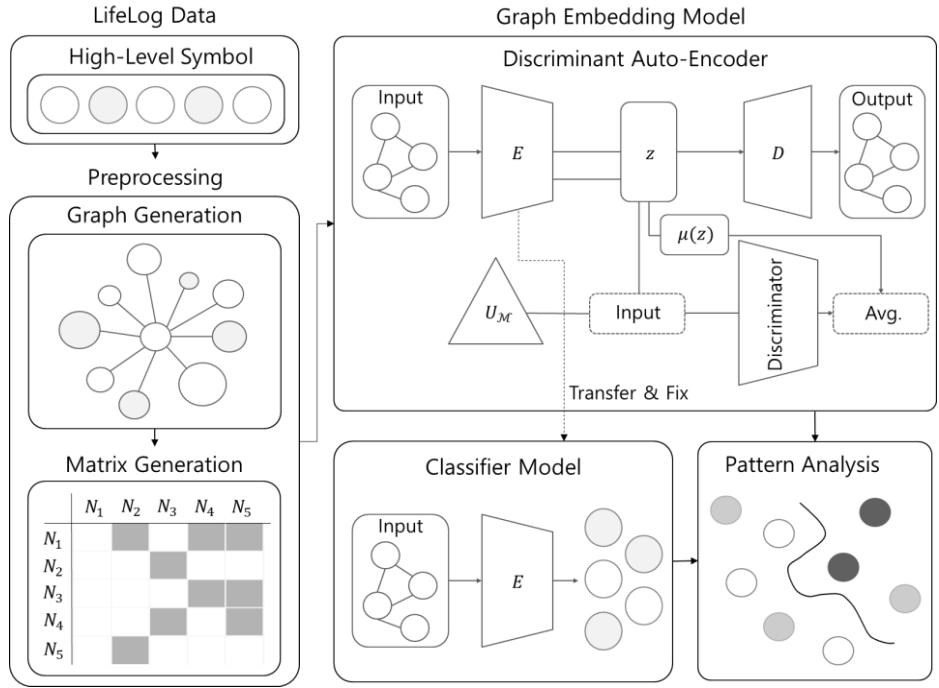

**Figure 2.** Proposed framework structure

**Table 2.** Semantics of network

| From | Semantic | To |
|---|---|---|
| sensor | contain | source |
| time | start | source |
| source | end | time |
| time | will be | time |

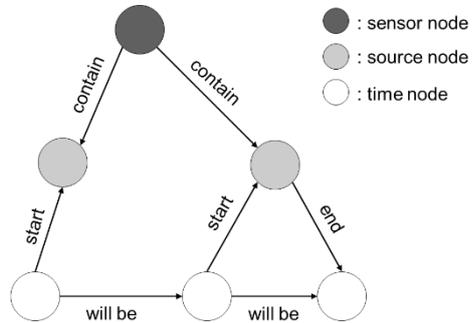

**Figure 3**. An example of network

### 3.3 Graph Embedding Process

We designed a graph embedding model inspired by [12]. [12] proposed the Adversarial Auto Encoder (AAE) model mapping latent space to Constant Curvature Manifold (CCM) M to improve the ability to express graph features in latent space. The CCM is defined as Eq. 1.

$$\mathcal{M} = \{x \in \mathbb{R}^{d+1} | <x,x> = \kappa^{-1}\} \tag{1}$$

$$<x,y> = x^T \begin{pmatrix} I_{d \times d} & 0 \\ 0 & -1 \end{pmatrix} y \tag{2}$$

The encoder of the proposed model uses the graph convolution layer [13]. Therefore, the input graph G is transformed according to the requirement of the layer. That is, the graph is represented by a node feature matrix $X$, which represents the characteristics of the node, and an adjacent matrix $A$, which represents the connection relationship between the nodes. $G = \{X, A\}$. Where $k$ means the size of the feature vector of the node and n is the number of nodes.

$$X \in \mathbb{R}^{n \times k} \tag{3}$$

$$A \in \mathbb{R}^{n \times n} = \begin{cases} m \text{ if } i \text{ and } j \text{ is } m \text{ times joined} \\ 0 \quad\quad\quad\quad else \end{cases} \tag{4}$$

The proposed model has the same learning mechanism as the general AAE [14]. Among them, the auto-encoder part receives $X$ and $A$ as inputs and outputs $X'$ and $A'$. In this way, the encoder learns the feature extraction ability for the graph. The discriminator induces the result of the encoder to exist in the CCM. Also, to improve the graph imbedding efficiency, the membership function $\mu(z)$ is added to the learning process of the discriminator. $\mu(z)$ has a value of 1 at the point of perfect existence in the CCM, but, converges to 0 at the opposite case. Therefore, it helps the discriminator to learn steadily. Finally, the encoder model maps the characteristics of the graph to latent space composed of CCM. Equations 5, 6, 7 and 8 show the object function of the auto-encoder, the object function of the discriminator, the encoder loss from discriminator, and the membership function, respectively. Where $\phi, \theta$ and $\lambda$ are weights of each models. And $\zeta$ is a hyperparameter that controls the width of the curve.

$$L_{AE}(\phi, \theta, X, A) = \mathbb{E}_{E_\phi(z|X,A)}[\log(D_\theta(X', A'|z))] \tag{5}$$

$$L_{Dis}(\phi, \lambda, X, A, z) = -\log Dis_\lambda(z) - \log\left(1 - Dis_\lambda\left(E_\phi(X, A)\right)\right) \tag{6}$$

$$L_{Enc}(\phi, \lambda, X, A, z) = -\log Dis_\lambda\left(E_\phi(X, A)\right) \tag{7}$$

$$\mu(z) = exp\left(\frac{-(<z,z> - \kappa^{-1})^2}{2\zeta^2}\right) \tag{8}$$

## 4 Experimental Result

### 4.1 UbiqLog (smartphone lifelogging) dataset

In this paper, we test the life log data provided by UCI machine learning repository [15]. The open source UbiqLog application is used to generate the UbiqLog dataset. This application is based on the participant's smartphone and has collected lifelogging data for at least two months from each participant. The data collection process was conducted in 2013 and 2014 for 35 participants. The UbiqLog dataset consists of seven sensors such as WiFi, Location, SMS, Call, Application Usage, Bluetooth Proximity, and Activity State and contains a total of 9,782,222 instances. All records contain high-level symbols and are human-readable. It has gender, user label.

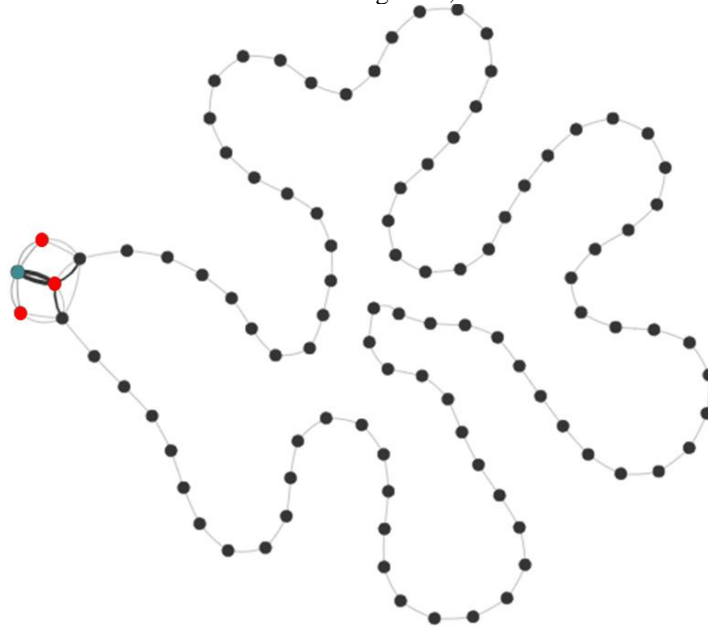

**Figure 4. graph visualization using JS**

### 4.2 Graph Visualization

To demonstrate that the data is well represented in the graph according to the designed semantics, we have created a graph visualization tool using Java Script (JS). Figure 4 shows an example plotting a graph using a visualization tool. In Figure 4, the black node represents the time node, the red node represents the source node, and the cyan node represents the sensor node. The thickness of the edge indicates the degree of connection. The proposed semantic network obtains time nodes every 15 minutes. Through this, it is possible to efficiently store an event occurring during a day using only a small memory space. It also provides additional information to the user's behavioral pattern analysis by placing a weight on the edge.

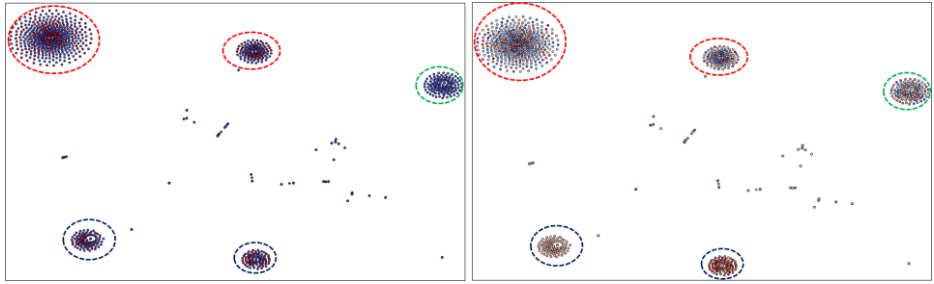

**Figure 5. t-SNE result of latent space with different label (left: sex, right: index)**

**Table 3. Classification accuracy with fixed encoder**

| Label | AE | CCM-AAE |
|---|---|---|
| Sex (2 class) | 70.65% ± 2.56% | **78.05% ± 1.1%** |
| Index (32 class) | 5.50% ± 1.20% | **15.42% ± 2.43%** |

### 4.3 Pattern Analyzation

In this section, we analyze the pattern of users in the dataset using the latent space formed by the learned encoder. For latent space analysis, we plotted the t-SNE for the latent variable. The two t-SNEs in Figure 5 are the same but reflect different labels (sex, human index). As a result, users' behaviors were classified into five patterns. In addition, these patterns are subdivided into patterns that most users have in common (red circle), patterns that are revealed in specific genders (green circle), and patterns that are revealed in specific users (blue circle). These results suggest that the proposed framework can be used in various ways. For example, a pattern corresponding to each cluster may be analyzed to provide a customized service targeting a specific user, a specific gender, or general users. However, the analysis of the pattern corresponding to each cluster was excluded because it did not fit the scope of this study.

Finally, we measure the accuracy by creating a deep classifier using fixed encoders. As shown in Table 3, the method used in this study showed better classification performance than the conventional AE. This indicates that CCM-AAE, a model used in our framework, can represent better graph features than basic AE.

## 5 Conclusion and Future work

In this paper, we propose a graph pattern analyzation framework for lifelog data. This framework can generate graphs that efficiently store daily lifelogs and embed the generated graphs to organize users' patterns. It also facilitates lifelog analysis through latent space analysis.

However, the daily graph can reduce the degree of information intensification due to severe fluctuations of size and inevitably mixed noises. Therefore, as a future study, we plan to study a framework that can extract subgraphs of daily graph and embed subgraphs to latent space.